# Transmission anomalies in 2D photonic crystals from the checkerboard family: From broken to hidden symmetries and plasmon "spoofing"

T. Dodge, M. Schiller, R. Yi, M.J. Naughton, and K. Kempa

Department of Physics, Boston College, Chestnut Hill, Massachusetts 02467, USA

**Abstract**

In a field representation, the main symmetry of the electromagnetic response of complementary metal film structures is described by the Babinet principle, expected to be obeyed by structures in vanishingly thin films of a perfect electric conductor. A softer transmittance Babinet principle (TBP) is not so restrictive. The goal of this work is to study how severely this broken symmetry affects the optical response of such structures. We consider two geometrically distinct series of planar complementary structures from the checkerboard family: regular and bowtie. The self-complementary structure of these series is very singular and breaks even the rigorous Babinet principle. We study complete simulated transmittance spectral maps ($T$-Maps) that accumulate the whole spectral response of an entire series of structures in a single plot. The *ab initio* $T$-Maps of these 2D photonic crystals were simulated for linearly polarized waves propagating perpendicular to the planar structures, made in a vanishingly thin film of a perfect electric conductor. While confirming the expected long wavelength validity of the TBP, we show that in the frequency range where diffraction effects dominate, the standard derivation of the TBP no longer applies, and with the help of our $T$-Maps, we demonstrate a total collapse of the TBP in the structures considered. This broken symmetry practically eliminates all but one transmission band on the hole side of the $T$-Maps, the remaining strong band being a "spoof" plasmon, free of multiple frequency replicas, an important feature for filter applications. By symmetry arguments and simulations, we discovered that the $T$-Maps for bowtie and doubled-period regular structures are identical. We discuss how this hidden symmetry can benefit applications by providing a convenient scaling, whereby simplified structures can deliver a tailored response.

Regular complementary structures [1] can originate from the familiar checkerboard pattern (Fig. 1, central pattern), consisting of a planar arrangement of squares (black, representing a metal in the physical analysis further below), point-touching at each corner. A series of related structures is obtained by uniformly scaling the dimensions of the black squares, while keeping the square-to-square center distance constant (equal to $a$). These structures consist of a square periodic arrangement of square islands of size $l_i = sa/2$ for scaling factors $s < 1$ (left panel of Fig. 1), and complementary arrays of square holes of size $l_h = (2 - s)\,a/2$ for $s > 1$ (right panel of Fig. 1). For the self-complementary checkerboard structure (center panel), $s = 1$, and so $l_i = l_h = a/2$. Note that the scaling factors of the corresponding complementary structures add up to 2.

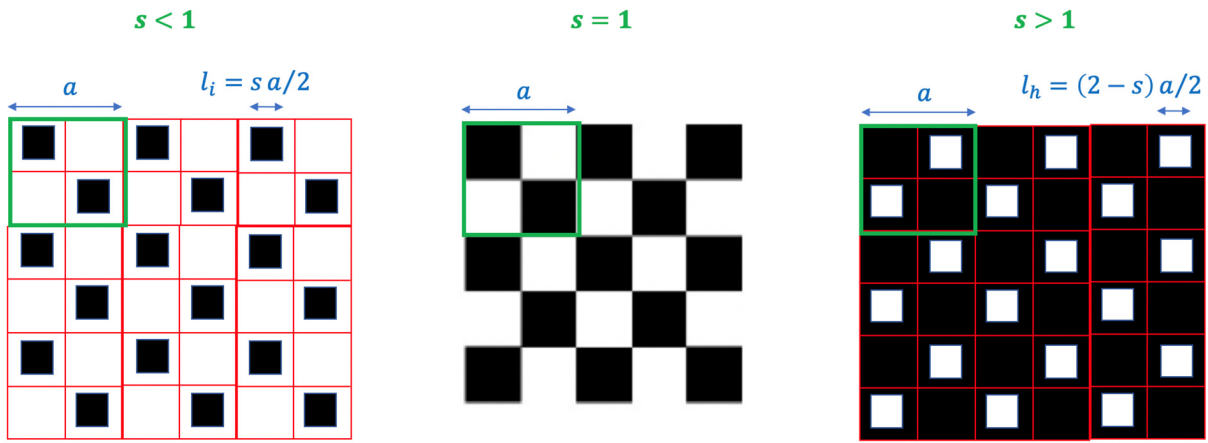

**Fig. 1**. The regular checkerboard series of structures: the perfect checkerboard (center panel), and it's complementary pairs (side panels). Green squares in each panel show the corresponding unit cells. The series is generated by uniformly scaling the island sizes while keeping their centers fixed.

Another important checkerboard series also originates from the familiar checkerboard, and is generated also by uniformly scaling the square sizes, but this time their central touching points are preserved. This is called a bowtie series, because the appearance of the resulting complementary structures.

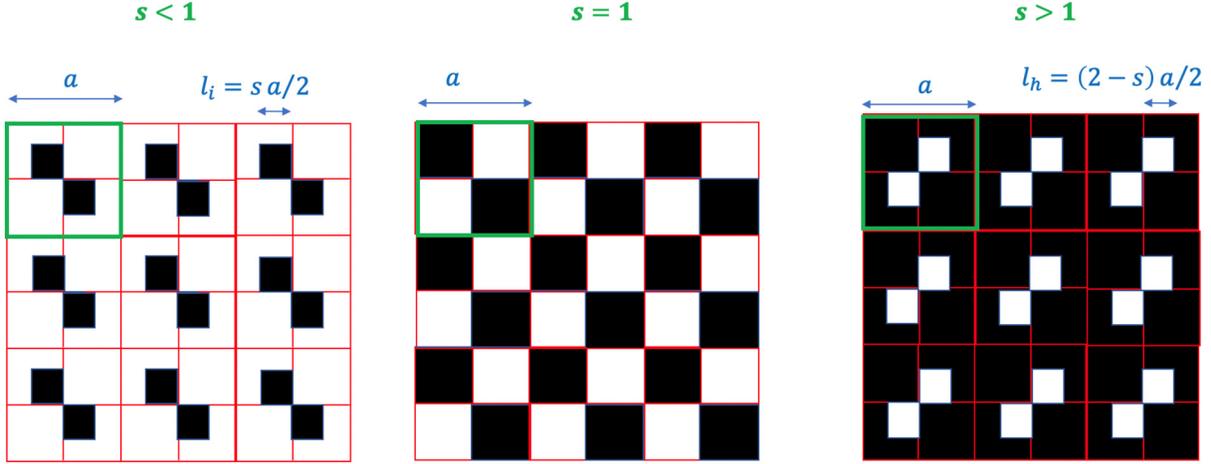

**Fig. 2**. The bowtie checkerboard series of structures: the perfect checkerboard (center panel), and its complementary pairs (side panels). Green squares in each panel show corresponding unit cells. The series is generated by uniformly scaling the island sizes while preserving their central touching point in each unit cell.

The electromagnetic response of both series of structures (for $s \neq 1$) made of vanishingly thin perfect electrical conductors (PEC) obeys the Babinet principle (BP), which has a rigorous form in the field representation [2-5] (see also Eq. 2, below). A more practical transmission representation of the BP, hereafter the TBP, commonly used in metamaterial optics, is [6,7]

$$T_1 + T_2 = 1 \qquad (1)$$

where $T_1$ is the transmittance of a structure and $T_2$ that of the corresponding complementary structure. The standard derivation of Eq. (1) in our context goes as follows. A structure for $s < 1$ (left panel in Fig. 1 or 2) lies in *x-y* plane. An initial transverse harmonic plane wave is propagating along *z*, and its fields are $\mathbf{E}_1^i = (E_1^i, 0, 0)$ and $\mathbf{H}_1^i = (0, H_1^i, 0)$. We assume no diffraction, and therefore a transmitted wave, sufficiently far from $z = 0$, also propagates along *z*, and has fields $\mathbf{E}_1 = (E_1, 0, 0)$ and $\mathbf{H}_1 = (0, H_1, 0)$. After replacing this structure with the complementary structure for $s \to 2 - s$, for which the complementary $s > 1$ (right panel in Fig. 1 or 2), the 90º rotation can be compensated by rotation of the initial wave, with fields $\mathbf{E}_2^i = \eta \mathbf{H}_1^i$ and $\mathbf{H}_2^i =$ $-\mathbf{E}_1^i/\eta$, where $\eta$ is the wave impedance [3]. The fields of the transmitted wave from the rigorous BP in the field representation are [3]

$$\mathbf{E}_2 = \eta(\mathbf{H}_1^i - \mathbf{E}_1) \qquad (2a)$$

$$\mathbf{H}_2 = (\mathbf{E}_1 - \mathbf{E}_1^i)/\eta \qquad (2b)$$

and the corresponding Poynting vector is $\mathbf{S}_2 = \mathbf{E}_2 \times \mathbf{H}_2 = (\mathbf{H}_1^i - \mathbf{H}_1) \times (\mathbf{E}_1 - \mathbf{E}_1^i)$. For $t = \mathbf{E}_1/\mathbf{E}_1^i = \mathbf{H}_1/\mathbf{H}_1^i$, and $\mathbf{S}_1^i = \mathbf{E}_1^i \times \mathbf{H}_1^i = \mathbf{E}_2^i \times \mathbf{H}_2^i = \mathbf{S}_2^i$, the transmittance $T_2 = \mathbf{S}_2/\mathbf{S}_2^i = (1 - t)^2 = r^2 = R_1 = (1 - T_1)$, which completes the proof of Eq. (1). Because of the no-diffraction assumption used in this derivation, Eq. (1), and therefore the TBP, has limited validity.

To study the structures from both regular and bowtie series, we performed numerical simulations of the transmittance. Our main simulations were implemented *via* the finite integration technique (FIT) algorithm on the CST Studio Suite commercial EM solver [8]. Since all of our structures are made of vanishingly thin PECs, these are essentially *ab initio* simulations, as only geometry enters the simulations of the Maxwell's equations, with the PEC entering for boundary conditions. Checkerboard structures were modeled to be periodic arrays from a single unit cell of PEC and vacuum geometries, under boundary conditions enforcing normal incidence of a TEM mode, with fields along square sides. Figure 3(b) shows a color-coded plot of the so-simulated $T$ as a function of $s$ and $f$ (frequency), the $T$-Map, done here for a transverse, linearly polarized wave propagating perpendicular to structures from the *regular* checkerboard series, with period $a = 4\ \mu m$.

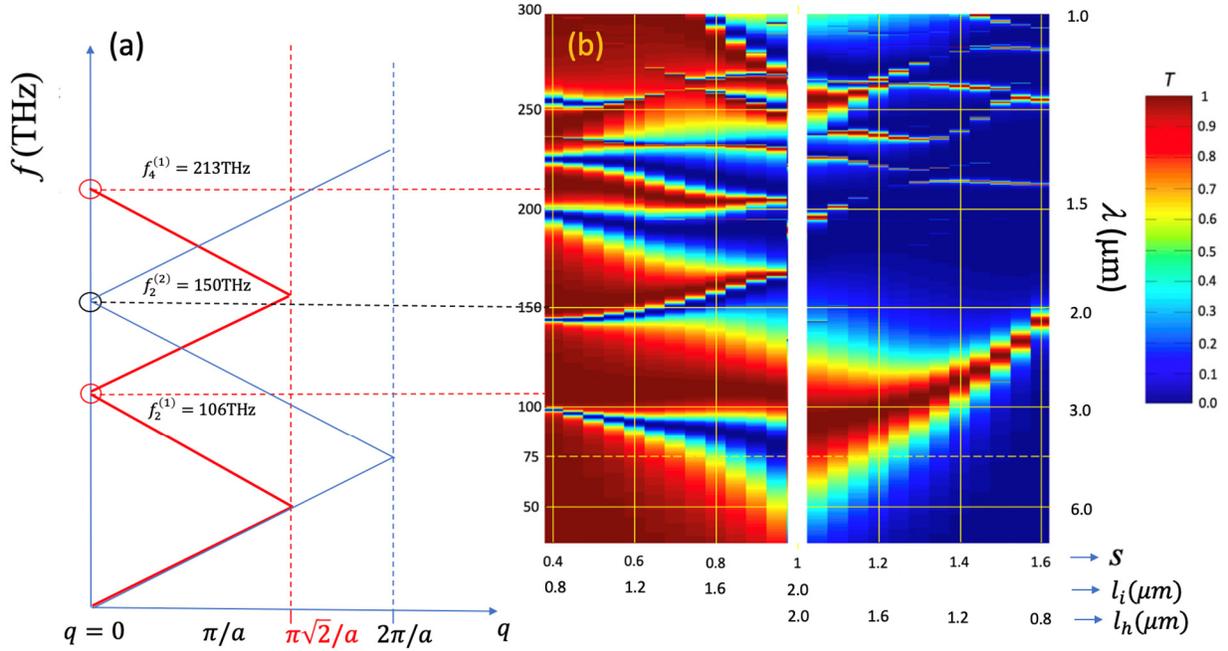

**Fig. 3**. (a) Sketch of the selected (dominant) 2D photonic bands on the island side of the regular checkerboard structure with $a = 4\ \mu m$ and $s \to 0$. The lowest three, energy-momentum conserving excitations occur in our case at the frequencies mark with circles. (b) $T$ vs. $s$ and frequency $f$ (the $T$-Map) simulated for the regular checkerboard series of structures with period $a = 4\ \mu m$. All structures are made of an infinitely thin PEC, exposed to an electromagnetic wave at normal incidence, with electric field linearly polarized parallel to the square edges.

The accuracy of our simulations has been established in a series of additional tests, involving an alternative code, COMSOL [9]. Figure 4 presents an example of such a comparative study, by showing simulations of $T$ spectra for (a) a regular structure with $s = 0.6$ and period $a$, and (b) its complement with $s = 1.4$ and the same period. Solid lines represent simulations via CST, and dashed lines via COMSOL, with structures in various polarization configurations. Figure 4 confirms that the results are not affected by enforcing the TE and/or TM modes of the EM wave at almost perpendicular propagation to the plane of the structures, as well as by an in-plane rotation of the structures by $\pi/4$ (rotated structure configuration). Additional confirmation of results was performed using other solvers available on the CST Studio Suite, including the finite element method (FEM) and the transmission-line matrix (TLM) method. There is overall excellent agreement, giving a large degree of confidence to the applied simulation codes.

The response of a perfect checkerboard structure ($s = 1$) has not been simulated, as this a very special, and highly singular, case discussed in detail elsewhere [1, 10]. In the long wavelength limit, dielectric function analysis leads to the simple result $T = 1/2$, a constant that is independent of frequency and period/square size. Confirming this simple result via simulations or experimental has proven to be a challenge [10]. The main reason is the fact that this self-complementary structure violates the BP, as demonstrated and discussed in detail in Ref. [10]. In addition, numerical simulations typically fail, since these employ iterative algorithms that in the case of this very singular structure are exponentially sensitive to initial conditions (Lyapunov exponent), which leads to chaos and numerical bifurcations, akin to the classic logistic map problem [1]. A truly perfect checkerboard structure is essentially impossible to realize in practice, since the slightest imperfection triggers bifurcations. It was demonstrated, however, that a slightly modified checkerboard with resistive (lossy) contacts placed between the corners of the touching squares removes this singular behavior, at least for sufficiently low frequencies [10]. Such a modified structure recovers, at least approximately, the expected nearly constant, broadband response. In radio technology, this broadband response has been known for over half a century [11], and became a basis for broadband (*e.g.* bowtie [12]) radio antennas.

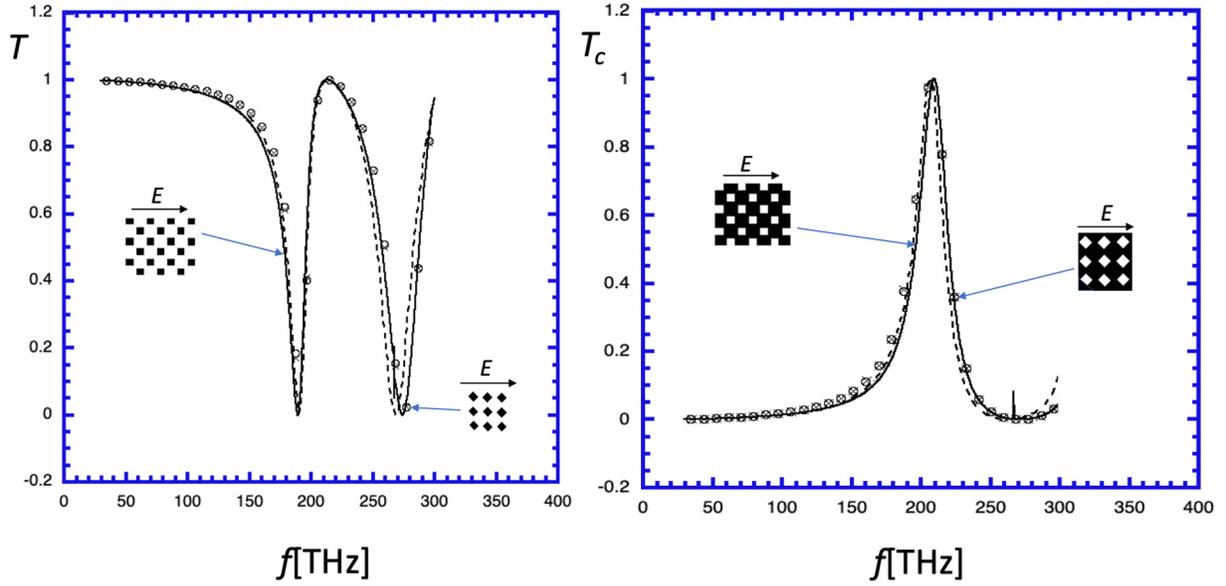

**Fig. 4**. Simulations of $T$ and $T_c$ vs. $f$, for two complementary structures $s = 0.6$ (a) and $s = 1.4$ (b). Structures are made of PEC with $d = 0$ and $a = 2\ \mu m$. Solid lines represent simulations via CST, and dashed lines via COMSOL, with structures in the normal configuration and polarization (see sketches on the left in both panels). Symbols are for CST simulation for $\pi/4$ rotated structures (see sketches on the right): circles are for the TM mode enforcement at near vertical propagation, and crosses are for the TEM mode enforcement.

The color $T$-Map of Fig. 3(b) reveals the *2D photonic crystal* nature [13] of the square island arrays of regular structures for $s < 1$. It consists of bands of transmission gaps (blue) in an otherwise highly transmitting background (red). The gaps narrow with reducing $s$, and for $s \to 0$, the structures reduce to square arrays of increasingly well-separated and vanishingly small conducting square islands. These can be viewed in a 45° rotated frame as arrays of diamond-like islands, with lattice period $\bar{a} = a/\sqrt{2}$ (see right inset in left panel of Fig. 4). The zero$^{th}$ order model of such a photonic crystal is a square array of conducting points, and the resulting non-interacting 2D photonic band structure is sketched in Fig. 3(a). Since the first BZ of such a 2D photonic structure is a square of $2\pi/\bar{a}$ size (along the edge of the square), the main symmetric directions in momentum space, expressed in terms of $a$, are: $0 < q < \pi\sqrt{2}/a = (\Gamma - K\ \text{direction})$ and $0 < q <$ $2\pi/a$ ($\Gamma - L$ direction). Assuming no electromagnetic coupling between the points, the free space photon dispersions (light lines) along these lines can be obtained from the free space photon dispersion $\omega = qc$ along these high symmetry (square island geometry) directions, but Umklapped at each BZ edge. Since our initial wave is assumed to propagate perpendicular to the 2D photonic crystal surface, the projected (into the 2D crystal plane) component of its wavevector is zero. Therefore, only crossings of the $q = 0$ line in Fig. 3(a) with the Umklapped photon dispersions represent scattering which conserves energy and momentum. These are given by

$$\lambda_n^{(i)} = a\, f(i)/n \tag{2}$$

where $n = 2, 4, ...,$ $f(1) = \sqrt{2}$, and $f(2) = 1$. For the three lowest crossings, Eq. (2) gives: $\lambda_4^{(1)} = 1.41\ \mu m$, $\lambda_2^{(2)} = 2\ \mu m$, and $\lambda_2^{(1)} = 2.83\ \mu m$. The corresponding frequencies are $f_4^{(1)} = 213$ THz, $f_2^{(2)} = 150$ THz, $f_2^{(1)} = 106$ THz. At these points, the incoming wave couples into a mode of the 2D photonic crystal, and thus is trapped in the 2D plane and does not contribute to $T$. As evident from Fig. 3, this zero$^{th}$ order analysis predicts rather well the locations of the first three transmittance suppression bands. The observed and expected broadening of these bands for larger values of $s$ results from the increasingly important interactions between the islands due to reduced inter-island distance.

Another feature observed in the $T$-Map in Fig. 3(b) is the overall failure of the TBP. Equation (1) implies that, in order for the TBP to be satisfied, the maps must be antisymmetric to a color change about the $s = 1$ column (from blue to red, and *vice versa*), i.e. $T_1(s) = 1 - T_2(2-s)$, for $s \neq 1$. As expected (see derivation above), this is the case only for sufficiently small frequencies, at which no diffraction occurs, indicated by appearance of the photonic bands or gaps. In Fig. 3(b), this onset occurs for $f < 50$ THz. Otherwise, the TBP fails completely, with suppressed transmission on the hole side, with only one, well defined transmission band dominating the *T*-Map. This *broken TBP symmetry* determines the physics of the hole side structures.

Since the periodic arrangement of holes is a legitimate 2D photonic crystal, one might expect to have a similar Umklapp band formation as on the island side. However, the basic mode in the vanishing hole size limit is not a free photon, but a photon that "fits" into the hole, *i.e.* it is diffraction limited to wavelengths sufficiently smaller than the hole size. In a simple model, the hole side of the complementary pair of the checkerboard structure represents an array of identical, vanishingly short square waveguides that cannot support transverse EM (*i.e.*, TEM) modes [3,4]. Thus, transmission for the wave propagating perpendicular to the screen must occur above the cut-off waveguide frequencies $f_h$. In the simplest case of a single waveguide, the fundamental TE$_{10}$ mode (the lowest frequency) is given by [3]

$$\lambda_h = c/f_h = 2l_h \tag{8}$$

This formula must be taken with a grain of salt, since the waveguides are vanishingly short. Note that this formula is strongly violated in the observed strong transmission band. Since our structures are made of PEC, no plasmonic effects can explain this unexpected transmission band. However, this mode can be explained by the "spoof" plasmon concept of Pendry [14], developed to explain the extraordinary optical transmission (EOT) [15] of light through subwavelength holes in structures similar to ours. By employing a similar, local waveguide approach, this theory [14] demonstrates that the photon field in such arrays "spoofs" plasmon-like behavior. The in-plane dielectric function of the effective uniform film (PEC + holes) can be defined in the long wavelength limit, and it acquires a Drude form $\varepsilon \sim 1 - \frac{\omega_p^2}{\omega^2}$, with the "spoofed" plasmon frequency given by $\omega_p = 2\pi f_h$, and $f_h$ given by Eq. (8). The corresponding dispersion of the spoofed plasmon is similar to surface plasmons, with the main branch dispersing along the light line for small $k_{||}$ (in-plane wavevector) and then asymptotically approaching $f_h$ for increasing $k_{||}$. Thus, the spoof plasmon, which enables transmission through the effective film, has its frequency always smaller (red-shifted).

To investigate this effect in our structures, we simulated a modified $T$-Map, showing $T$ vs. $a$ and $f$ for the hole structures with fixed $l_h = 1\ \mu m$, made of PEC of vanishing thickness. Again, the spectrum is dominated by a strong transmission band near 120 THz. This strong "spoof" plasmon resonance is essentially independent of $a$ for large $a$. However, its frequency is red-shifted (by ~ 20%) below the value expected from Eq. (8), $f_h = 150$ THz, in agreement with the "spoof" theory of Pendry [14].

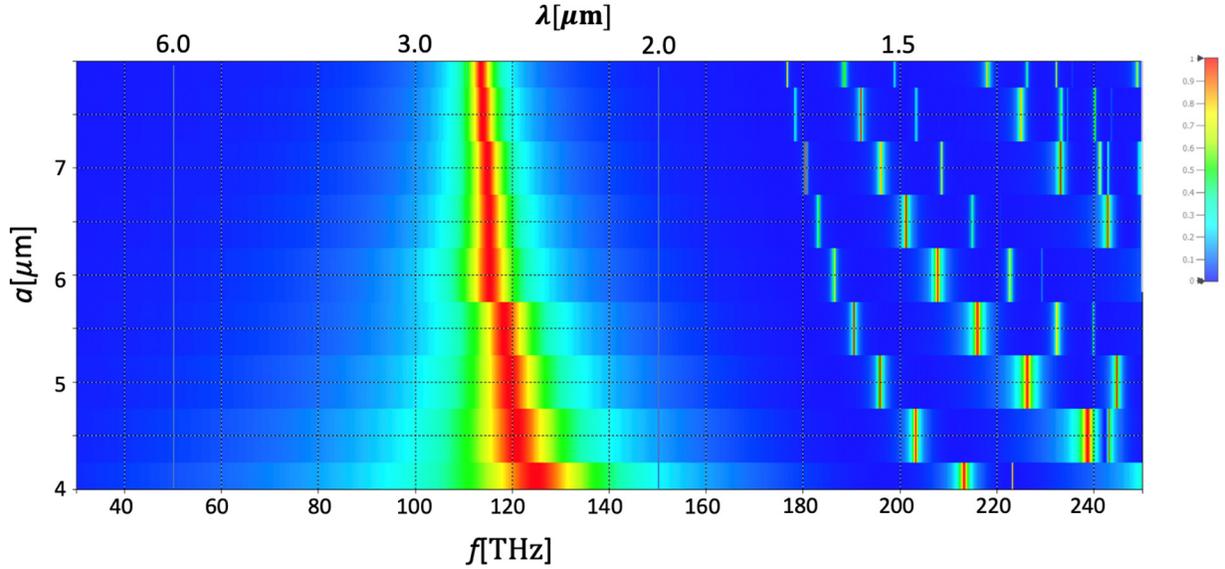

**Fig. 5**. Modified $T$-Map of $T$ vs $a$ and $f$ for the hole-side checkerboard structures with fixed $l_h = 1\ \mu m$, made of PEC of vanishing thickness.

The most dramatic, *hidden symmetry* in the family of the checkerboard structures can be identified by comparing regular to bowtie versions of these structures. Fig. 6 illustrates this by comparing side-by-side $T$-maps for a regular checkerboard series (with $a = 4\ \mu m$), and for a bowtie series (with a period of $a = 2\ \mu m$). Striking, and unexpected, is the fact that these *T*-Maps are *identical*, apart from barely visible numerical inaccuracies.

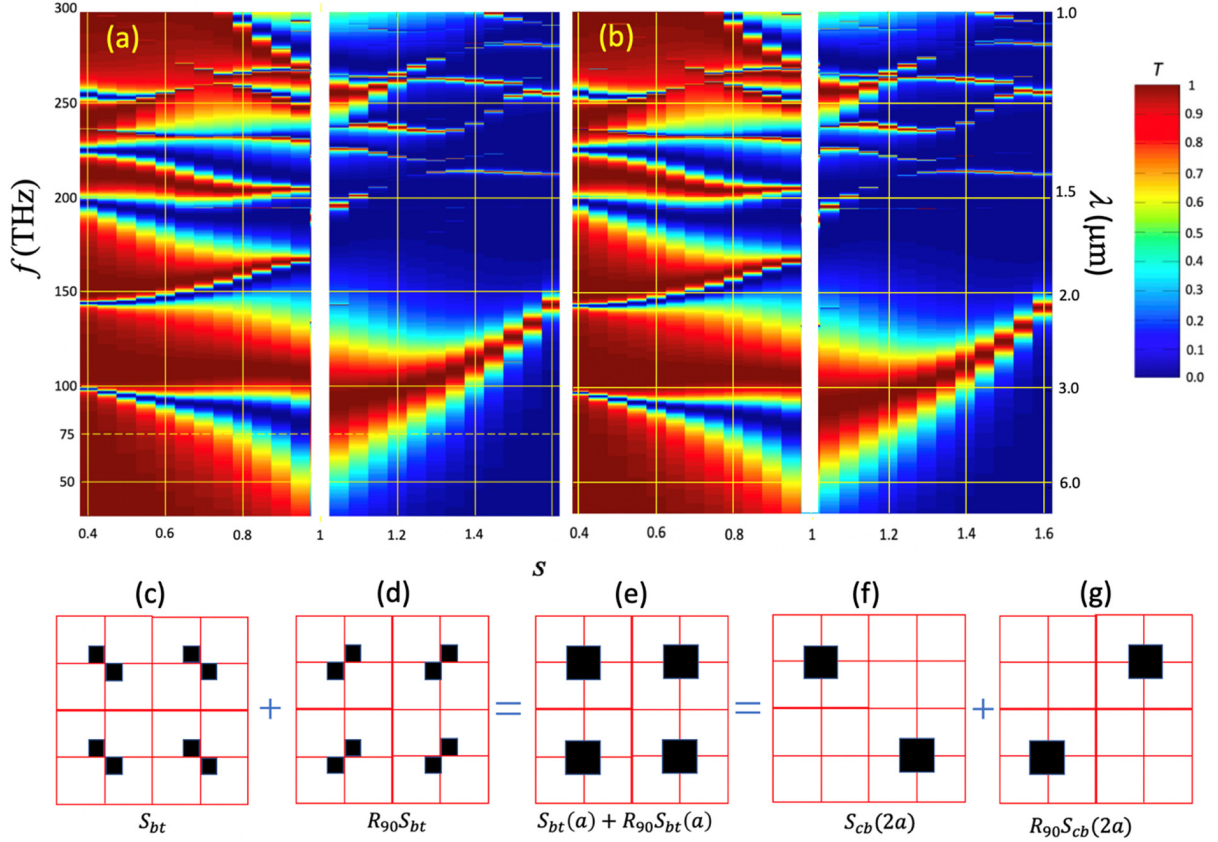

**Fig. 5.** *T*-maps for: (a) *regular* structure series with $a = 4\ \mu m$, and (b) *bowtie* series with $a = 2\ \mu m$. In (c – d) modifications of the bowtie structures into the regular structures of double period. Details in text.

To justify this result, recall that our structures are 2D photonic crystals made of PEC elements, and as such, they can be broken down conceptually into elemental units of perfectly lossless point dipoles. In addition, the excitation of the structures is by a linearly polarized plane wave propagating perpendicular to the surface of the structures, *i.e.* all elemental point dipoles are excited in-phase. Also, we are interested in total transmittance that is normalized in-plane and time averages of the Pointing vector. Keeping these facts in mind, we begin by analyzing the geometric relations between regular and bowtie structures (on the island side, $s < 1$) shown in Fig. 5(c-g).

While the bowtie structure of period $a$ (Fig. 5c) is marked $S_{bt}$, the 90° rotated bowtie structure of the same period is $R_{90}S_{bt}$ (Fig. 5d), where $R_{90}$ is the rotation operator. Clearly, by adding the two bowtie structures, we arrive at Fig. 5(e), which in turn is simply a sum of two regular checkerboard structures (Fig. 5(f) and 5(g), the last one 90° rotated), but with doubled period $2a$ (see Fig. 5). Based on the analysis above and Figs. 5 (c-g), the following relations are expected for the reflectance $R$:

$$R[S_{bt}(a)] = R[R_{90}S_{bt}(a)] \qquad (3)$$

$$R[S_{cb}(2a)] = R[R_{90}S_{cb}(2a)] \qquad (4)$$

$$R[S_{bt}(a) + R_{90}S_{bt}(a)] = R[S_{bt}(a)] + R[R_{90}S_{bt}(a)] \qquad (5)$$

Equations (3) and (4) simply reflect the obvious symmetry of the problems, and the *linearity* of the reflective response in Eq. (5) is expected because of the arguments at the beginning of this paragraph. Using Eq. (4) and an analog of Eq. (5) for regular checkerboard series structures, we can write:

$$R[S_{cb}(2a) + R_{90}S_{cb}(2a)] = R[S_{cb}(2a)] + R[R_{90}S_{cb}(2a)]$$
$$= 2R[S_{cb}(2a)] = R[S_{bt}(a) + R_{90}S_{bt}(a)] = 2R[S_{bt}(a)] \qquad (6)$$

Because the structures are lossless, $R + T = 1$, we finally get the result demonstrated via simulations in Fig. 5(a) and 5(b):

$$T[S_{cb}(2a)] = T[S_{bt}(a)] \qquad (7)$$

Similar arguments can be used on the hole side of the *T*-map (for $s > 1$), and the same result will be obtained. It is clear that our justification of this result relies on the linearity of Eq. (5) and its analog for the regular structures. Our conjecture is that such reflectance or transmittance linearity is universal, and can be used in the analysis of *T*-Maps for all checkerboard structures. While this analysis is for the selected checkerboard series (bowtie and regular), other checkerboard series, or in general other complementary series, should have a similar hidden symmetries, and the discovered linearity of the *R* (or *T*) responses might also hold.

Finally, we note that both the broken TBP symmetry and the hidden symmetry between regular and bowtie structures could be used in applications. The broken TBP symmetry and the resulting "spoof" transmission band feature on the hole side of the *T*-Map can be significant for bandpass filter applications that typically suffer from multiple harmonic replicas of the response, with multiple pass-bands at higher (and undesirable) frequencies. Clearly, the hole side of our *T*-Maps is free of that; there is a negligible transmission activity only at higher frequencies for structures with $s > 1.4$. Similarly, the unusual hidden symmetry between the checkerboard and bowtie structures might also lead to applications, by suggesting various replacement strategies between various bowtie and regular structures.

In conclusion, *ab initio* simulated *T*-Maps for 2D photonic crystals formed by structure series belonging to the checkerboard family reveal that the transmission version of the Babinet principle is not valid in frequency regions occupied by photonic bands. The immediate consequence of this is *broken symmetry* in the *T*-Map transmittance spectra representation between the island and hole sides, with a dramatic reduction of the number, width and strengths of the transmission bands on the hole side. In fact only one, strong "spoof" plasmon band dominates the spectrum there, free of harmonic replicas, an important feature for applications. Our systematic study of the *T*-Maps, combined with the symmetry based, comparative analysis of the bowtie *vs*. regular structures, reveals *hidden symmetry* that leads to the identity of the *T*-response between a bowtie structure and the regular structure of doubled period. We expect this last discovery to lead to other hidden symmetries in a broader class of complementary structures.